\documentclass[twocolumn]{aastex701}

\newcommand{\GW}{GW25}
\newcommand{\CAND}{3,554}
\newcommand{\fraccand}{3.6\%}
\newcommand{\obscand}{5,550}
\newcommand{\turnon}{40\%}
\newcommand{\turnoff}{19\%}
\newcommand{\intermediate}{41\%}
\newcommand{\newcand}{3,481}

\usepackage{booktabs}
\usepackage{amsmath}
\usepackage{subcaption}
\usepackage{hyperref}
\usepackage{graphicx}
\usepackage{txfonts}
\usepackage{lipsum}
\usepackage{subcaption} \usepackage{ulem}
\usepackage{cancel}
\usepackage{booktabs}
\usepackage{tabularx}

\begin{document}

\title{CLAS+: A large catalog of Changing-Look AGN candidates selected through S-PLUS narrow-band photometry}

\author[orcid=0000-0001-6480-1155]{L. Nakazono}
\affiliation{Observat\'orio Nacional/MCTI, Rio de Janeiro, RJ, Brazil}
\affiliation{Departamento de Física Matemática, Instituto de Física, Universidade de São Paulo,
R. do Matão 1371, 05508-090, São Paulo, SP, Brazil}
\email[show]{liliannenakazono@on.br}

\author[]{L. L. Oliveira}
\affiliation{Instituto de Astronomia, Geof\'isica e Ci\^encias Atmosf\'ericas, Universidade de S\~ao Paulo, S\~ao Paulo, SP, Brazil}
\email[]{leticialanzadeoliveira@gmail.com}

\author[orcid=0000-0001-8847-0047]{R. R. Valen\c{c}a}
\affiliation{Instituto de Astronomia, Geof\'isica e Ci\^encias Atmosf\'ericas, Universidade de S\~ao Paulo, S\~ao Paulo, SP, Brazil}
\email[]{raquelrvalenca@gmail.com}

\author[orcid=0000-0003-0820-4692]{P. Sánchez-Sáez}
\affiliation{European Southern Observatory (ESO), Karl-Schwarzschild-Str. 2, 85748 Garching bei München, Germany}
\email[]{}

\author[orcid=0009-0003-6609-1582]{G. Oliveira-Schwarz}
\affiliation{Escola Politécnica, Universidade de São Paulo, São Paulo, 05508-010, SP, Brasil}
\email{}

\author[orcid=0000-0002-2238-9665]{N. M. Cardoso}
\affiliation{Escola Politécnica, Universidade de São Paulo, São Paulo, 05508-010, SP, Brasil}
\email{natanael.mc@usp.br}

\author[orcid=0000-0002-6656-5333]{A. L. O'Mill}
\affiliation{CONICET. Instituto de Astronomía Teórica y Experimental (IATE), Laprida 854, Córdoba X5000BGR, Argentina.}
\affiliation{Universidad Nacional de Córdoba (UNC). Observatorio Astronómico de Córdoba (OAC), Laprida 854, Córdoba X5000BGR, Argentina}
\email{}

\author[orcid=0000-0002-6656-5333]{S. Panda}
\affiliation{International Gemini Observatory/NSF NOIRLab, Casilla 603, La Serena, Chile}
\email{}

\author[orcid=0000-0002-8280-4445]{E. Telles}
\affiliation{Observatório Nacional / MCTIC, Rua General José Cristino 77, Rio de Janeiro, RJ, 20921-400, Brazil}
\email{}

\author[orcid=0000-0003-3921-2177]{R. Demarco}
\affiliation{Institute of Astrophysics, Facultad de Ciencias Exactas, Universidad Andr\'es Bello, Sede Concepci\'on, Talcahuano, Chile}
\email{}

\author[orcid=0000-0002-4064-7234]{W. Schoenell}
\affiliation{The Observatories of the Carnegie Institution for Science, 813 Santa Barbara St, Pasadena, CA 91101, USA}
\email{}

\author[orcid=0000-0002-0138-1365]{T. Ribeiro}
\affiliation{Rubin Observatory Project Office, Tucson, AZ 85719, USA}
\email{}

\author[orcid=0009-0007-8005-4541]{A. Kanaan}
\affiliation{Departamento de Física, Universidade Federal de Santa Catarina, Florianópolis, 88040-900, SC, Brazil}
\email{}

\author[orcid=0000-0002-7736-4297]{C. Mendes de Oliveira}
\affiliation{Instituto de Astronomia, Geof\'isica e Ci\^encias Atmosf\'ericas, Universidade de S\~ao Paulo, S\~ao Paulo, SP, Brazil}
\email{}

\collaboration{all}{The S-PLUS collaboration}

\begin{abstract}
 Changing-look active galactic nuclei (CLAGN) exhibit rapid spectral transitions on timescales of months to years, challenging standard AGN unification models and providing unique insight into accretion-disk variability and obscuration processes. Despite their scientific importance, systematic searches for CLAGN remain limited by the scarcity of multi-epoch spectroscopy and the difficulty of constructing large samples efficiently.
   We introduce CLAS+, a catalog constructed by comparing synthetic photometry derived from DESI DR1 and SDSS DR17 spectra with contemporaneous narrow-band imaging from the S-PLUS survey. 
   Applying our pipeline to a parent sample of 98,139 quasars, we select high-confidence CLAGN candidates above a threshold of reduced chi-squared $\chi_r^2 > 15$.
   We identify \CAND{} strong CLAGN candidates, corresponding to \fraccand{} of the parent sample. This selected fraction is conditional on the adopted criterion and should not be interpreted as an intrinsic CLAGN occurrence rate. Cross-matching with existing CLAGN compilations shows that CLAS+ identifies a large population of previously unreported high-priority candidates (\newcand{}). Among the spectroscopic--photometric comparisons with reliable temporal classification, \turnon{} are classified as photometric turn-on, \intermediate{} as intermediate, and \turnoff{} as turn-off. We also find an apparent redshift dependence in the relative fractions, with photometric turn-on candidates becoming more common at higher redshifts within the CLAS+ selected sample.
   Therefore, narrow-band photometry provides a powerful and efficient alternative to purely spectroscopic searches for identifying CLAGN candidates at scale. CLAS+ substantially expands the known CLAGN candidate population and provides a valuable target list for future spectroscopic follow-up and time-domain studies with facilities such as Rubin/LSST.
\end{abstract}


\section{Introduction}
\label{sec:intro}
Changing-look active galactic nuclei (CLAGN) are characterized by dramatic spectral transitions in which broad emission lines appear or disappear, typically accompanied by strong variability in the blue featureless continuum across optical, ultraviolet, and X-ray bands. These transitions occur over a wide range of rest-frame timescales, from decades down to months or even days, challenging standard thin-accretion-disk models. CLAGN is primarily an observational classification, while the underlying physical drivers are commonly grouped into two broad categories: changing-obscuration AGN (COAGN), where clouds or outflows temporarily obscure the central engine, and changing-state AGN (CSAGN), which arise from intrinsic changes in the accretion flow onto the central supermassive black hole. Proposed intrinsic mechanisms include accretion-disk instabilities, structural disk transitions, evolution of the X-ray corona, and magnetically arrested disk states, while extrinsic scenarios involve variable obscuration, tidal disruption events, and other nuclear transients. In most observed cases, however, the transitions are consistent with changes in the accretion state that modify the ionizing continuum and, consequently, the structure and visibility of the broad-line region (see review by \citealt{2023NatAs...7.1282R}).

The first changing-look events were reported nearly five decades ago by \citet{1976ApJ...210L.117T}. Initially identified through only a handful of individual discoveries (e.g., \citealt{1995ApJ...443..617S}; \citealt{2014ApJ...788...48S}; \citealt{2014ApJ...796..134D}; \citealt{2015ApJ...800..144L}), the field has rapidly evolved in recent years, transitioning toward increasingly large and systematically identified samples due to large spectroscopic surveys and time-domain searches (e.g., \citealt{2018ApJ...862..109Y}; \citealt{2019ApJ...874....8M}; \citealt{2021AA...650A..33P}; \citealt{2022ApJ...933..180G}; \citealt{2023MNRAS.524..188L}; 
\citealt{2023MNRAS.518.2938T}; \citealt{2024ApJ...966...85Z}; \citealt{2025ApJS..278...28G}), with new CLAGN candidates being reported at an accelerating pace each year.

Most previous searches for CLAGN have relied on broad-band photometric variability, spectroscopic changes, or a combination of photometric pre-selection and spectroscopic follow-up. Large-amplitude variability thresholds, for example $|\Delta g| > 1$ mag, have been used to identify changing-look quasar candidates for follow-up spectroscopy (e.g., \citealt{2016MNRAS.457..389M}; \citealt{2019ApJ...874....8M}), while other searches have selected CLAGN through spectral transitions directly (e.g., \citealt{2018ApJ...862..109Y}).

In this work we present CLAS+, a systematic search for changing-look AGN candidates based on narrow-band photometry from the Southern Photometric Local Universe Survey (S-PLUS; \citealt{2019MNRAS.489..241M}) and synthetic photometry derived from spectroscopic observations from the Dark Energy Spectroscopic Instrument (DESI; \citealt{2016arXiv161100036D}, \citealt{2026AJ....171..285D}), complemented by the (extended) Baryon Oscillation Spectroscopic Survey (BOSS/eBOSS; \citealt{2013AJ....145...10D}, and \citealt{2016AJ....151...44D} ) from the Sloan Digital Sky Survey (SDSS; \citealt{2000AJ....120.1579Y}). By comparing spectroscopically-derived pseudo-photometry with observed S-PLUS measurements, we identify sources showing significant discrepancies consistent with strong spectral variability. This approach takes advantage of the enhanced sensitivity of narrow-band photometry to localized spectral changes, particularly variations in strong emission lines, and therefore our search is restricted to redshift intervals in which a prominent quasar emission line falls within one of the S-PLUS narrow-band filters. The paper is organized as follows: Section \ref{sec:Photometric data} presents the data used in this work. Section \ref{sec:methods} describes the methodology. Section \ref{sec:results} presents the results, Section \ref{sec:catalog} describes the catalog, and Section \ref{sec:conclusions} summarizes our main findings.
\section{Photometric Data: S-PLUS}

\label{sec:Photometric data}

The Southern Photometric Local Universe Survey (S-PLUS) is a multi-band optical survey designed to map approximately 9,300 square degrees of the southern sky using a dedicated 0.8-meter robotic telescope at Cerro Tololo Inter-American Observatory (CTIO) in Chile. The survey employs the Javalambre photometric system \citep{cenarro+19} consisting of 12 filters -- five broad-band filters similar to those of the SDSS (u, g, r, i, z) and seven narrow-band filters strategically centered on key stellar features.
In this work, we use photometric data from the sixth data release\footnote{Available only for internal members at the time of paper submission. When the DR6 becomes public, the data will be available at \url{https://splus.cloud}} (DR6; Oliveira-Schwarz et al. in preparation), reduced by the Multiband Astronomical Reduction package \citep{mar}. We use the PSF aperture magnitudes, and the median limiting depth at a signal-to-noise ratio threshold of S/N $\geq 3$ are 21.6, 22.3, 22.1, 21.7, 21.2 for $u,g,r,i,z,$ respectively. The depths are 21.3, 20.8, 20.9, 21.0, 21.2, 21.8, 20.9 for the narrow bands J0378, J0395, J0410, J0430, J0515, J0660, J0861, respectively.

\begin{figure}
    \centering
    \includegraphics[width=1\linewidth]{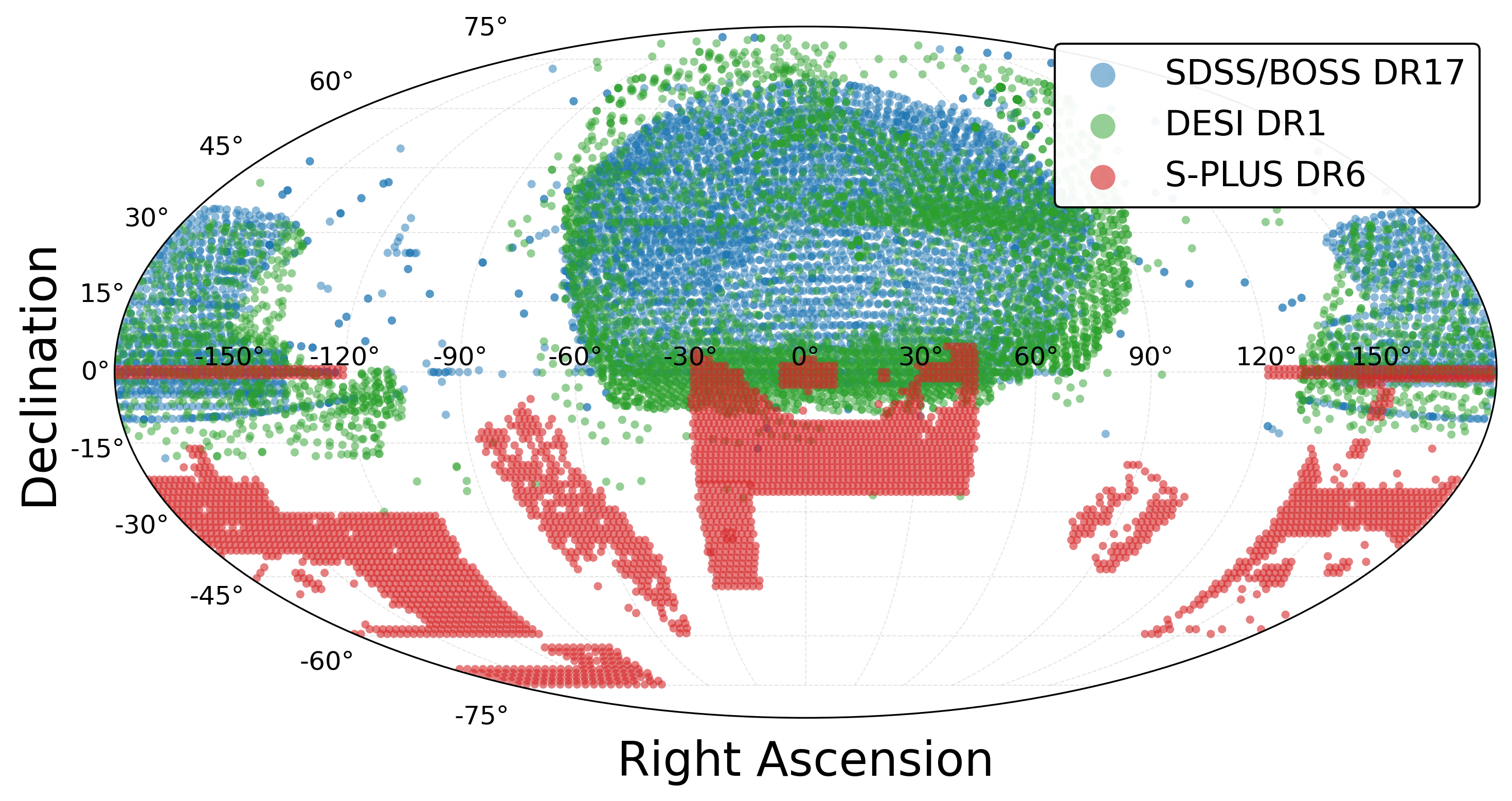}
    \caption{Footprint showing the overlapping regions of SDSS/BOSS DR17 plates (blue), DESI DR1 plates (green), and S-PLUS DR6 fields (red), mostly in the equatorial region.}   
    \label{fig:footprint}
\end{figure}

\section{Methods}
\label{sec:methods}
We compare synthetic photometry derived from DESI spectra with observed photometric measurements from S-PLUS, instead of comparing spectra obtained at different epochs. This approach allows us to exploit the much larger sky coverage and cadence of photometric surveys to identify potential CLAGN candidates.

\subsection{Selecting parent sample}
We use spectroscopic observations from DESI DR1 \citep{2025arXiv250314745D},
supplemented by SDSS/BOSS DR17 spectra \citep{2022ApJS..259...35A} when
available. DESI provides the primary spectroscopic reference sample for
this work because it contains the largest number of quasars overlapping
the S-PLUS footprint, while SDSS/BOSS provides an additional earlier
spectroscopic epoch for a subset of sources.

The pipeline was applied to 121,798 quasars crossmatched between S-PLUS and DESI through the AI-Scope\footnote{\url{https://ai-scope.cbpf.br/}} platform \citep{oliveira_schwarz_2026_19543949} and considering 1\arcsec{} crossmatch. Spectra were retrieved from SDSS and DESI through the \texttt{sparcl} package \citep{2025ASPC..541...77J}. In addition to the S-PLUS/DESI cross-matched sample, a total of 14,199 quasars in S-PLUS were only observed by SDSS DR17. Because DESI DR1 provides a substantially larger number of spectra than SDSS in our sample, we focus our methods on comparing S-PLUS with DESI. We use SDSS DR17 spectra to evaluate additional evidence of variability for the comparison between S-PLUS and DESI. 

Because the main objective of this work is to build a fully automated selection framework, we restricted the analysis to these surveys. 
The spatial distribution of the dataset is shown in the footprint of the Figure \ref{fig:footprint}. The SDSS I-IV spans between 2000--2020, DESI DR1 ranges between 2021--2022, whereas S-PLUS DR6 has observations from 2016--2023.


\subsection{Generating synthetic S-PLUS photometry from spectra}



When comparing spectroscopic observations to  photometry, we compute synthetic photometry by convolving the spectrum through filter transmission curves. Wavelength coverage is handled by masking out bad pixels so only valid spectral regions are used, then each filter’s transmission curve is interpolated onto that wavelength grid with out‑of‑range values set to zero. Therefore, the band integration only uses the overlap between the spectrum and the filter and does not extrapolate beyond the spectrum’s limits. \textbf{} The uncertainty in this synthetic photometry must account for both statistical spectral noise and systematic calibration uncertainties. For a spectrum with a flux density $f_\lambda(\lambda)$ and a filter with a transmission function $T(\lambda)$, the synthetic flux in a given band is computed as:

$$F_{\rm band} = \frac{\int f_\lambda(\lambda) \cdot T(\lambda) \cdot \lambda d\lambda}{\int T(\lambda) \cdot \lambda d\lambda}.$$
 
In practice, this integral is evaluated on the discrete wavelength grid of the spectrum $\{\lambda_i\}$ with pixel widths $\Delta\lambda_i$:

$$F_{\rm band} = \frac{\sum_i f_i \cdot T(\lambda_i) \cdot \lambda_i \cdot \Delta\lambda_i}{\sum_i T(\lambda_i) \cdot \lambda_i \cdot
  \Delta\lambda_i} = \sum_i w_i f_i,$$

where the  band-energy normalized weights are $$w_i = \frac{T(\lambda_i)\lambda_i\Delta\lambda_i}{\sum_j T(\lambda_j)\lambda_j\Delta\lambda_j}.$$

For the S-PLUS narrow bands (FWHM $\sim 15$–$30$ Å), the variation of $\lambda$ across the bandpass is small ($<1\%$) and has negligible impact. For the broad bands (u, g, r, i, z, FWHM $\sim 1000$ Å), the full wavelength dependence of this term is retained.

The statistical uncertainty propagated from the spectrum depends on the survey data products.

DESI provides a resolution matrix $R$ that encodes the pixel-to-pixel correlations introduced by the spectrograph's line-spread function. Including this correction, the propagated variance is

$$\sigma^2_{\rm stat} = \sum_i (R^T w)_i^2 \cdot \sigma^2_i,$$

\noindent
where $\sigma^2_i = 1/{\rm ivar}_i$. When the resolution correction is small, this reduces to the standard diagonal approximation

$$\sigma^2_{\rm stat} = \sum_i w_i^2 \sigma^2_i.$$

SDSS/BOSS spectra provide only diagonal inverse-variance arrays and do not include the full covariance structure introduced by resampling and sky subtraction. Previous studies have shown that the SDSS/BOSS pipeline uncertainties are mildly underestimated at the $\approx$10–20\% level depending on wavelength \citep[e.g.][]{2013AA...559A..85P}. To approximately account for these effects, we apply a conservative empirical variance inflation factor $\alpha=1.2$

$$\sigma^2_{\rm stat} = \alpha^2 \sum_i w_i^2 \sigma^2_i,$$

This correction should be regarded as an approximate empirical treatment rather than a first-principles covariance reconstruction.

Beyond statistical noise, synthetic photometry is subject to systematic uncertainties associated with filter transmission curves and spectrophotometric calibration. We adopt a conservative uncertainty floor of 3\% fractional uncertainty associated with filter throughput variations

$$\sigma_{\rm filt} = 0.03 \times \,|\,F_{\rm band}\,|\,.$$

This value is larger than the internal calibration scatter reported for S-PLUS DR5, for which the Gaia-based calibration yields millimagnitude-level residuals \citep{2026arXiv260715891V}, and is intended to account more broadly for possible uncertainties in the effective filter response and photometric calibration that are not captured by the quoted statistical errors.

We consider a conservative spectrophotometric calibration uncertainty of 5\%

$$\sigma_{\rm specphot} = 0.05 \times \,|\,F_{\rm band}\,|\,,$$
consistent with the few-percent calibration accuracy of SDSS/BOSS spectroscopy \citep{2013AJ....145...10D}.

The total per-band uncertainty is computed as

$$\sigma_{\rm total} = \sqrt{\sigma^2_{\rm stat} + \sigma^2_{\rm filt} + \sigma^2_{\rm specphot}}$$
We note that calibration uncertainties are partially correlated across filters. A full covariance treatment would therefore be required for precision SED fitting. However, because the objective of this work is candidate selection rather than parameter inference, we adopt this simplified independent-error approximation and empirically calibrate the variability threshold using the parent quasar population.

\subsection{Chi-squared calculation}
To identify highly variable quasars, we compare synthetic photometry derived from spectroscopy with observed S-PLUS photometry using the statistic as follows:

$$\chi^2_r = \frac{ \sum_n(F_{conv,n}-F_{obs,n})^2}  {\text{dof}(\sigma^2_{conv,n} + \sigma^2_{obs,n})},$$

where dof are degrees of freedom, $F_{conv,n}$ is the synthetic flux derived from the spectrum, $F_{obs,n}$ is the observed S-PLUS flux, and $\sigma_{conv,n}$ and $\sigma_{obs,n}$ are the corresponding uncertainties.

Because calibration uncertainties and spectrophotometric systematics are only approximately modeled, $\chi^2_r$ should not be interpreted as a formal statistical chi-square distribution. Instead, we use $\chi^2_r$ as an empirical variability metric quantifying the disagreement between spectroscopy-derived and observed photometry.

\subsection{Narrow-Band Emission-Line Masking}
\label{ssec:mask}

\begin{figure}[!ht]
\centering
\includegraphics[width=\columnwidth]
{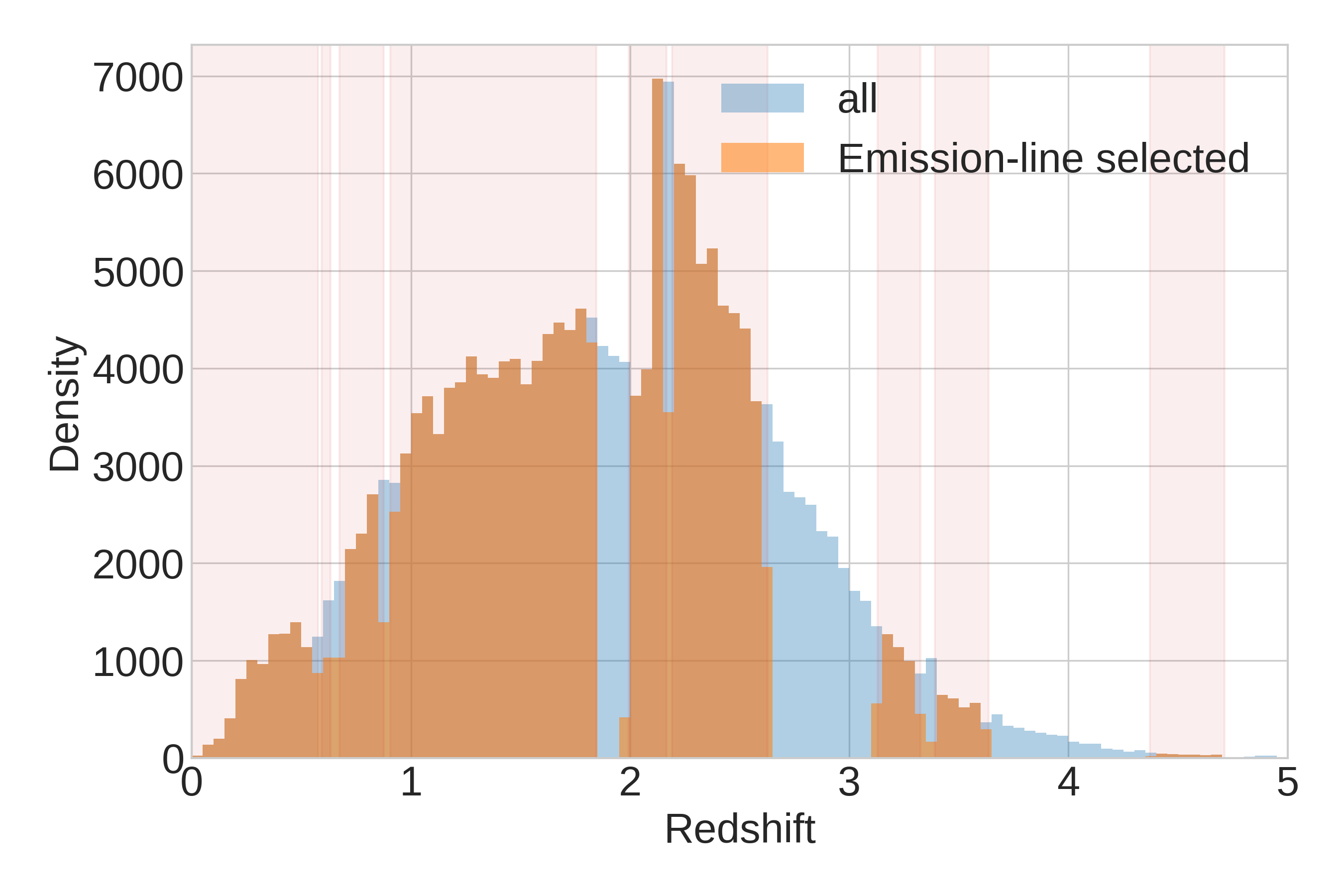}
  \caption{Redshift distribution of the full quasar sample (grey) and the emission-line selected subset (blue). Red shaded regions indicate the redshift windows where at least one quasar emission line falls within an S-PLUS narrow-band filter (threshold at 10\% of peak transmission). Objects outside these windows are excluded from the CLAGN search, as any excess in the narrow bands would not be attributable to emission-line variability.}
     \label{fig:redshift_window}
\end{figure}

To ensure that the $\chi^2_r$ excess detected in the S-PLUS narrow bands is physically meaningful, we restrict our sample to quasars whose known emission lines fall within the effective transmission window of at least one narrow-band filter. For each of the seven S-PLUS narrow bands (J0378, J0395, J0410, J0430, J0515, J0660, and J0861), we computed the effective wavelength interval where the filter transmission exceeds 10\% of its peak value, and then derived the corresponding redshift intervals in which fourteen common quasar emission lines (from Ly$\alpha$ at 1216 \AA to H$\alpha$ at 6565 \AA) would be observed within that band. 
The union of all per-line, per-filter redshift intervals defines a set of discrete redshift windows where a spectral feature is expected to contribute significantly to the narrow-band flux. Only objects whose spectroscopic redshift falls within at least one of these windows are retained for further analysis (see Figure \ref{fig:redshift_window}), resulting in a sample of 102,949 unique sources. This selection ensures that a genuine emission-line contribution is a plausible driver of the observed photometric $\chi^2_r$ excess between the S-PLUS narrow-band photometry and the flux predicted from the spectrum. 

\subsection{Avoiding contamination by the host galaxy or observational artifacts}

We investigated whether the observed variability is due to observational effects such as a nearby saturated star or poor photometry by checking the flagging from SExtractor\footnote{ \url{https://sextractor.readthedocs.io/en/latest/Flagging.html}} \citep{sextractor}. A total of 98,139 sources have an associated flag less than 2 and were kept in  our catalog for further analysis.  Figure \ref{fig:flag} shows the distribution of the reduced chi-squared for narrow- \textbf{($\chi^2_\text{r,narrow}$)} and broad-band fluxes \textbf{($\chi^2_\text{r,broad}$)}. The concentration around $\chi^2 \sim 1$ indicates that most of our parent sample have well-calibrated flux measurements. We, therefore, infer that the high-variability sources are unlikely to be driven by systematic effects in the imaging and that our selected CLAGN candidates represent genuinely high-variability sources according to our spectroscopic--photometric metric.


\subsection{Defining photometric turn-on/off and intermediate stages}
\label{ssec:turn}
The classification is based on the sign and magnitude of the flux residuals

$$    \Delta f_{\rm band} = f_{\rm spec,conv} - f_{\rm obs,SPLUS},$$
where $f_{\rm spec,conv}$ is the archival spectrum convolved through the S-PLUS filter 
transmission curves, and $f_{\rm obs,SPLUS}$ is the S-PLUS photometry. We define 
$t_1 < t_2 < t_3$ as the chronologically ordered observation epochs, where $t_i$ may 
correspond to either a spectroscopic or a photometric observation. The residual 
$\Delta f_{\rm band}$ is evaluated only in narrow band filters, and its sign reflects whether the 
source brightened or faded between the two epochs being compared.

\begin{itemize}
    \item A \textbf{photometric turn-off candidate} has $\langle\Delta f\rangle > 0$: 
    the source was brighter at the earlier epoch than at the later epoch, consistent 
    with fading broad-line emission over the interval $\Delta t = t_{i+1} - t_i > 0$.

    \item A \textbf{photometric turn-on candidate} has $\langle\Delta f\rangle < 0$: 
    the source was fainter at the earlier epoch than at the later epoch, consistent 
    with brightening broad-line emission over the same interval $\Delta t$.

    \item A \textbf{photometric intermediate candidate} is defined for sources with 
    three observation epochs $t_1 < t_2 < t_3$, where the residuals change sign 
    between consecutive pairs, i.e., 
    $\langle\Delta f(t_1, t_2)\rangle \cdot \langle\Delta f(t_2, t_3)\rangle < 0$. 
    This indicates that the source flux reversed its trend between the intervals 
    $[t_1, t_2]$ and $[t_2, t_3]$, consistent with a non-monotonic variability 
    event occurring within the total baseline $\Delta t = t_3 - t_1$.
\end{itemize}

This classification is photometric and does not follow the standard spectroscopic 
definition of CLAGN turn-on/turn-off, which requires the appearance or disappearance 
of broad emission lines confirmed in multi-epoch spectra. We therefore adopt the 
terms \textit{photometric turn-on candidate}, \textit{photometric turn-off candidate}, 
and \textit{photometric intermediate candidate} throughout this work.

\subsection{Selection pipeline}

The selection pipeline can be summarized as follows:
\begin{enumerate}
    \item Crossmatch the S-PLUS DR6 photometric catalog with the DESI DR1 quasar sample using a 1 arcsec radius and retrieve the spectra.
    \item Compute synthetic S-PLUS photometry by convolving each spectrum with the S-PLUS filter transmission curves.
    \item Compute $\chi^2_r$, $\chi^2_{r,\rm broad}$, and $\chi^2_{r,\rm narrow}$ by comparing the synthetic fluxes with the observed S-PLUS fluxes.
    \item Retain only sources for which at least one quasar emission line falls within an S-PLUS narrow-band filter.
    \item Remove sources with problematic SExtractor flags, adopting ${\tt flags\_det} > 2$ as the exclusion criterion.
    \item Select high-confidence CLAGN candidates using $\chi^2_{r,narrow} > 15$.
    \item For the selected candidates, repeat steps 2--6 using available SDSS/BOSS spectra within 1 arcsec, when present.
    \item Calculate photometric classification (turn-on, turn-off or intermediate)
\end{enumerate}

\section{Results}

\label{sec:results}

Applying the variability-selection pipeline to 121,798 quasars, we identify \obscand{} highly variable observations corresponding to \CAND{} unique candidate CLAGN systems (\fraccand{}). Multiple observations may exist for the same source because some objects are associated with more than one spectroscopic epoch.

\begin{figure}[!ht]
\centering
\includegraphics[width=\columnwidth]
{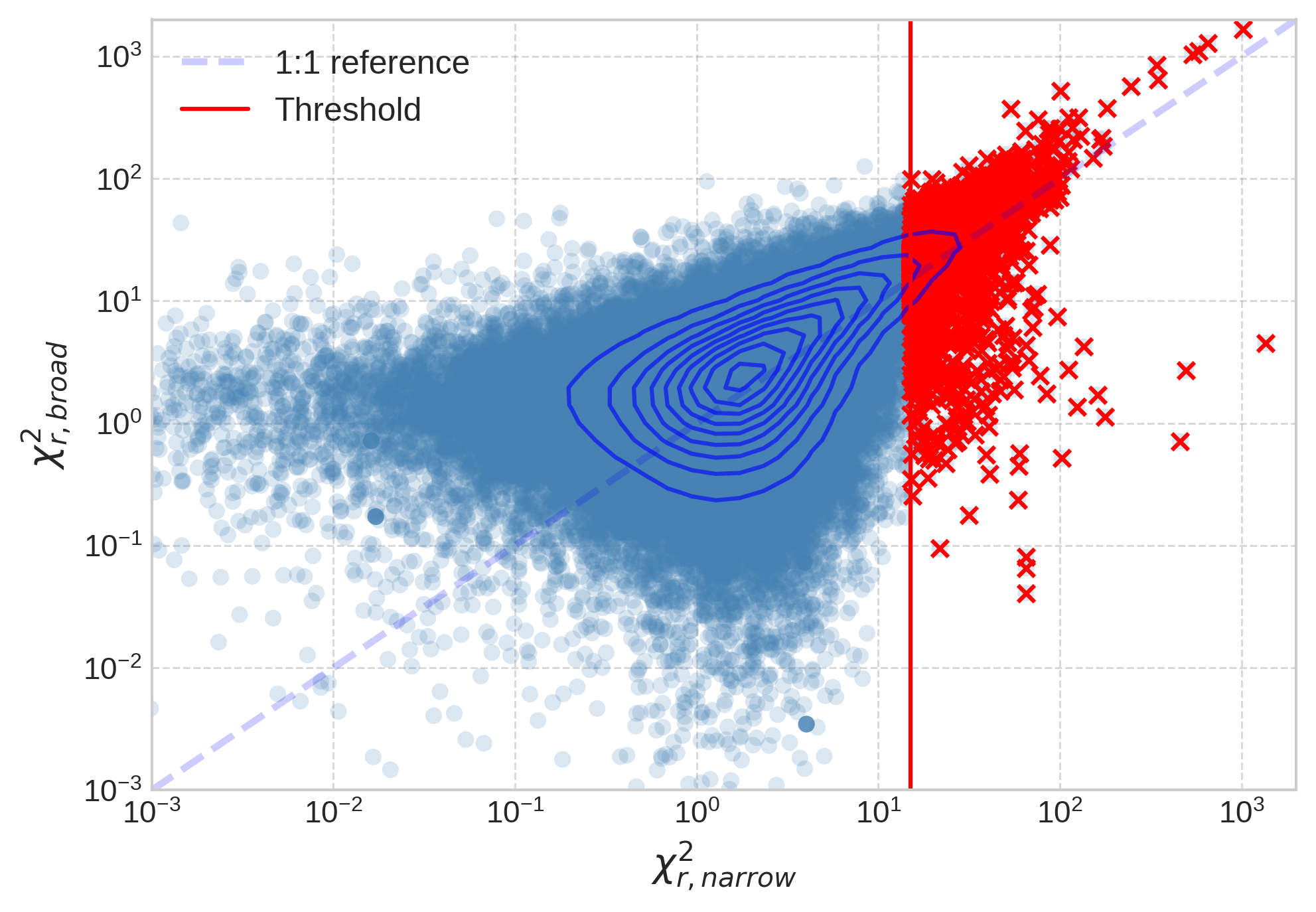}
  \caption{Distribution of reduced chi-squared calculated on broad versus narrow bands. The red vertical line represents the threshold adopted in this work to select CLAGN candidates.}
     \label{fig:flag}
\end{figure}

For that, we adopted $\chi^2_r>15$ as a conservative empirical threshold defining highly variable candidates. Most quasars cluster near $\chi^2_r\approx$  1–3, while a strong tail toward high $\chi^2_r$ values is present for extreme outliers (see Figure \ref{fig:flag}). We conservatively define candidate CLAGN as sources with $\chi^2_r > 15$, corresponding to the extreme tail of the empirical distribution. This threshold is intended to isolate the most statistically significant variability events and define a manageable sample of strong candidates for spectroscopic follow-up. The resulting selected fraction is therefore conditional on the adopted threshold and survey selection function and should not be interpreted as the intrinsic fraction of CLAGN in the parent population. A formal calibration of the threshold will require systematic follow-up of representative subsets of both selected and unselected sources. 

In the literature, previous studies of extreme AGN variability have adopted similar statistical thresholds to isolate genuine transitions (e.g., \citealt{2019ApJ...874....8M}; \citealt{2022ApJ...933..180G}). Relaxing this threshold to $\chi^2_r > 10$ would approximately double the catalog size (Table \ref{tab:chi2_thresholds}); however, it is not yet established whether the additional sources represent genuine but lower-amplitude CLAGN transitions or an increasing fraction of continuum variability unrelated to CLAGN-type transitions. Conversely, a stricter threshold ($\chi^2_r > 25$) isolates a smaller, higher-confidence subset, but we cannot currently rule out that it also excludes genuine CLAGN with more modest narrow-band excess. Discriminating between these possibilities requires spectroscopic follow-up across the threshold range (Sect. \ref{sec:conclusions}).

\begin{figure}[!ht]
\centering

\begin{subfigure}{\columnwidth}
    \centering
    \includegraphics[width=\linewidth]{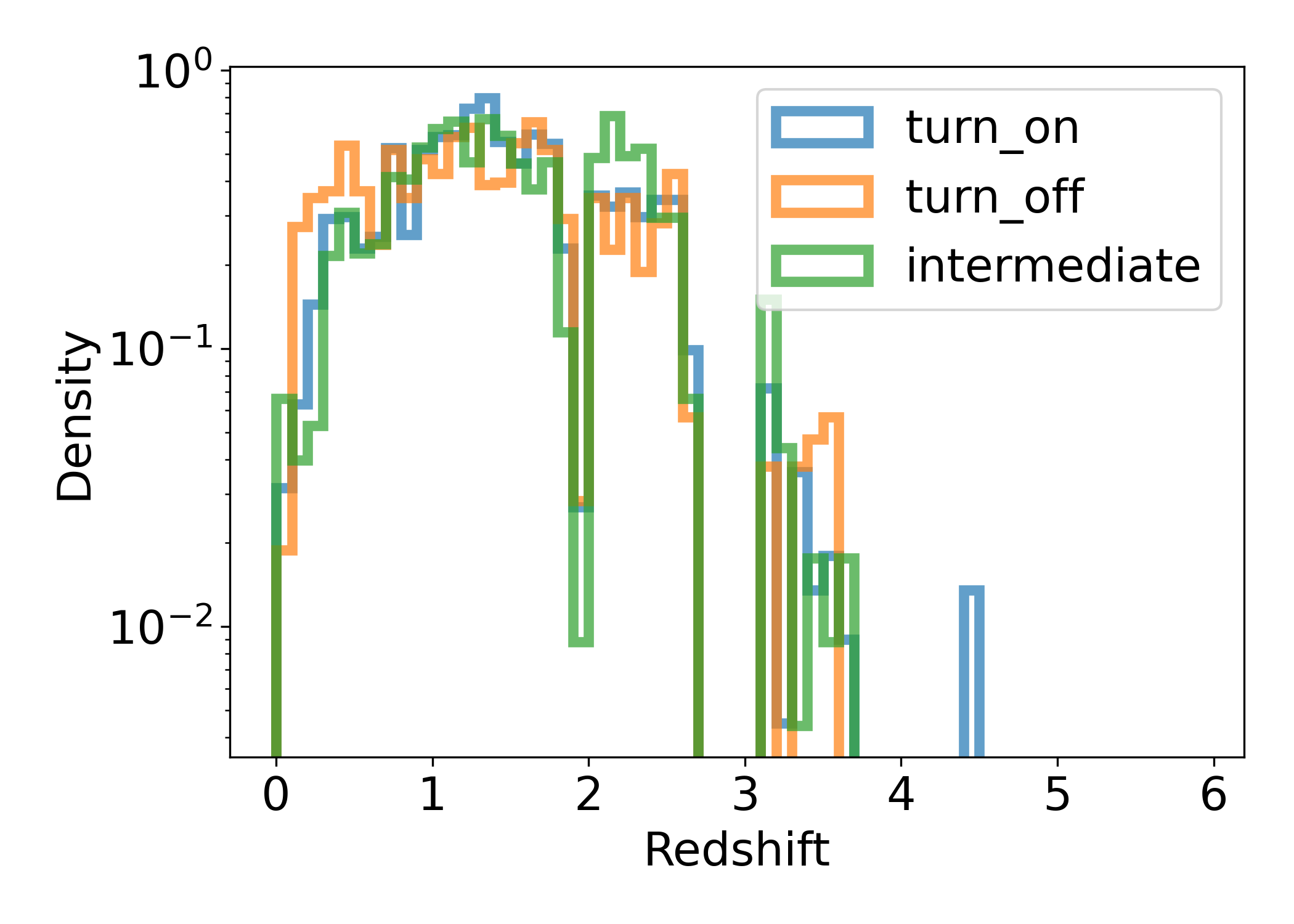}
    \caption{Redshift distribution}
    \label{fig:redshift}
\end{subfigure}

\vspace{0.3cm}

\begin{subfigure}{\columnwidth}
    \centering
    \includegraphics[width=\linewidth]{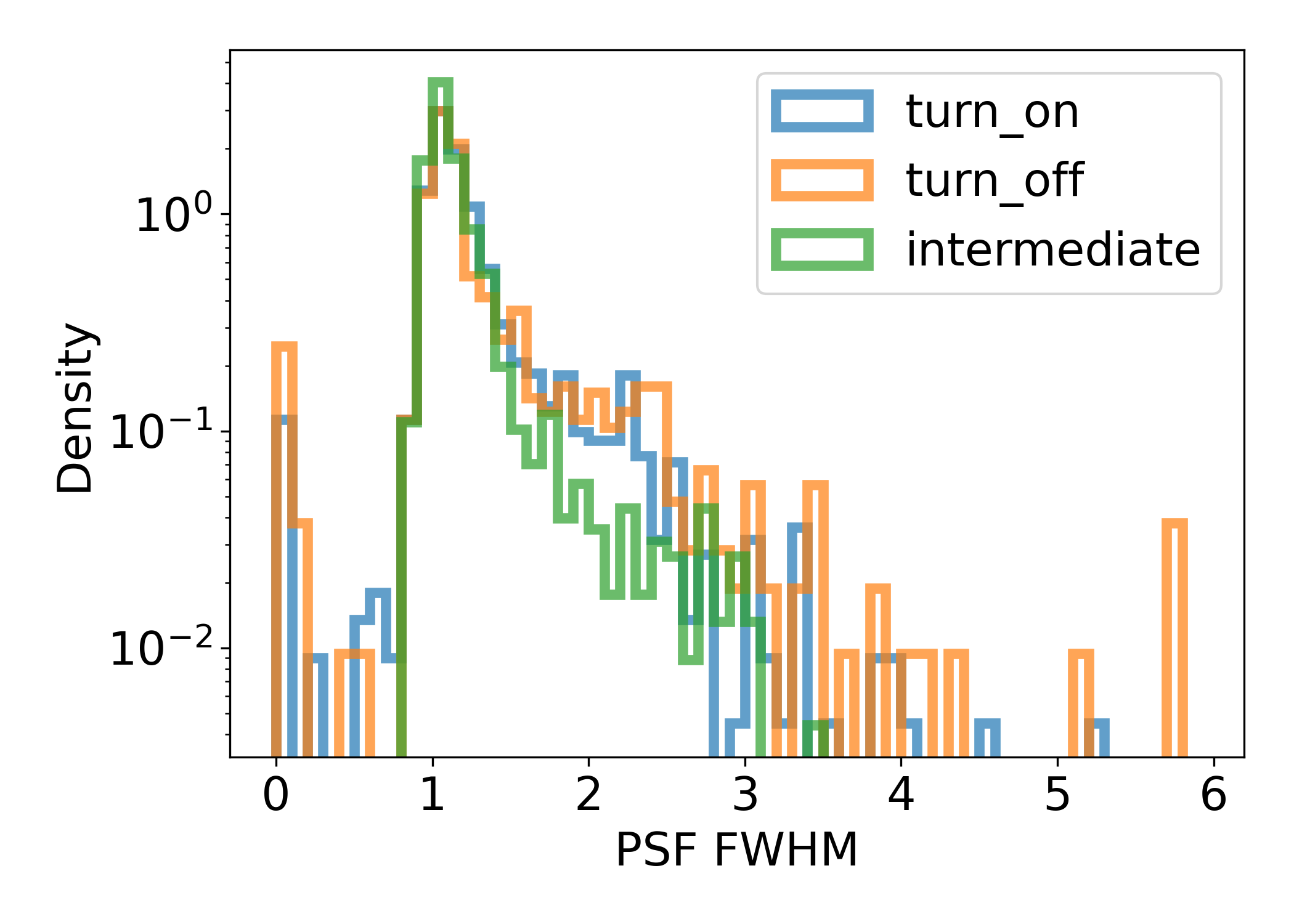}
    \caption{Normalized PSF FWHM distribution}
    \label{fig:fwhm}
\end{subfigure}

\begin{subfigure}{\columnwidth}
    \centering
    \includegraphics[width=\linewidth]{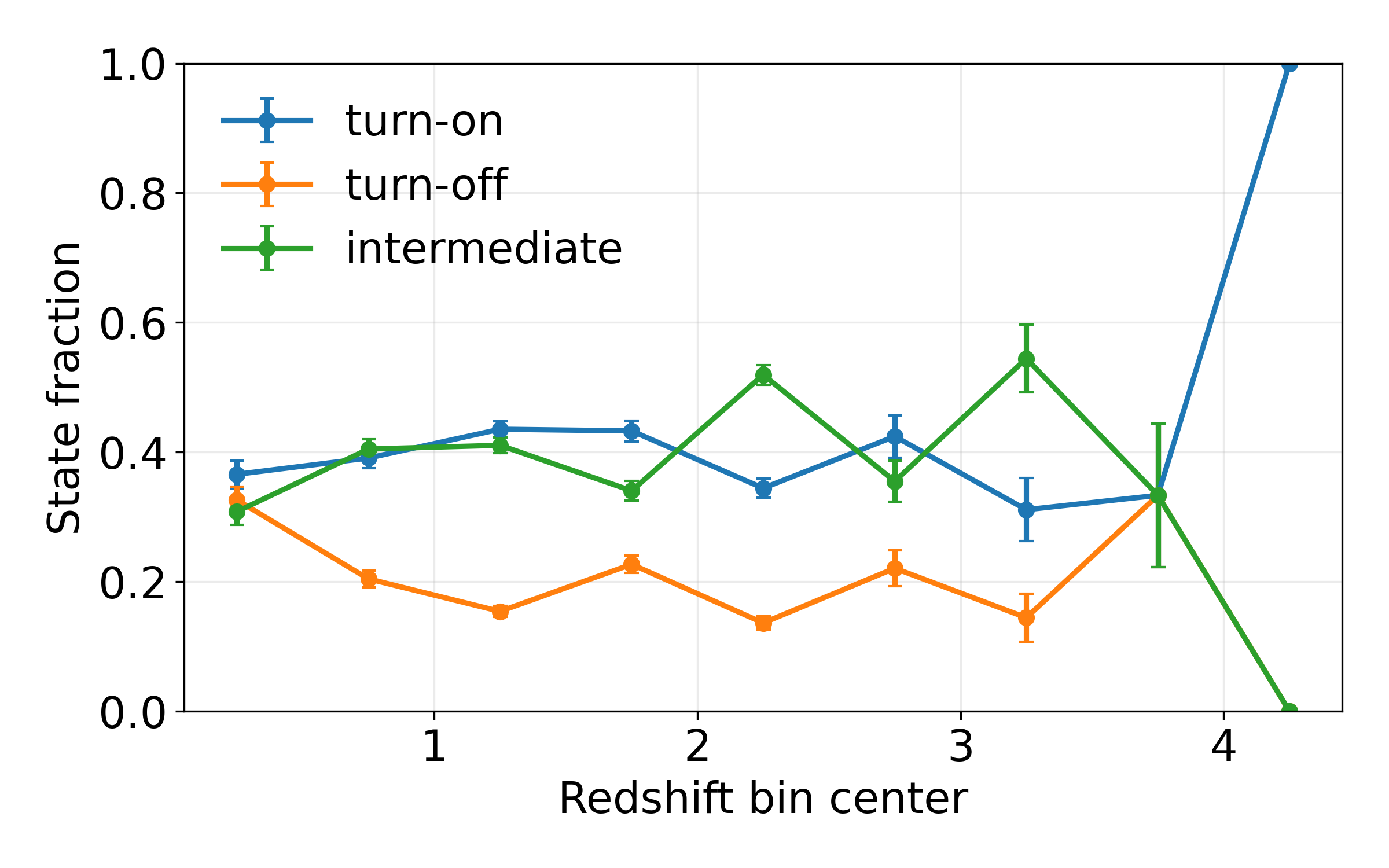}
\caption{The fractions of turn-on and turn-off CLAGN candidates are similar at low redshifts, but there is a higher prevalence of turn-on at high redshifts. Error bars indicate binomial uncertainties, $\sigma_f = [f(1-f)/N]^{1/2}$, where $f$ is the class fraction and $N$ is the number of classified comparisons in each redshift bin.}
\label{fig:redshift_state}
\end{subfigure}

\caption{Distributions for photometric turn-on, turn-off and intermediate states.}
\label{fig:combined}
\end{figure}

 To ensure that our catalog is dominated by point-like objects, we inspected their full width at half maximum (FWHM) of the point-spread function. The normalized PSF FWHM is defined as the ratio between the measured source FWHM and the median PSF FWHM of the corresponding S-PLUS field; values close to unity indicate unresolved sources. As shown in Figure \ref{fig:fwhm}, most selected sources are unresolved, suggesting that host-galaxy contamination is not a dominant effect in our analysis. In addition, our initial sample includes only sources classified as quasars, of which most are $z > 1$ (Fig. \ref{fig:redshift}).

By requiring high-quality prior spectroscopic observations and setting a conservative variability threshold, we are likely missing lower-amplitude CLAGN events or those with more gradual transitions (also seen in Figure \ref{fig:dist}). The redshift distributions of the three photometric classes are broadly similar (Fig.~\ref{fig:fwhm}). This suggests that the classification is not dominated by a gross redshift imbalance among the classes. However, the detectability of each state can still depend on the CLAS+ selection function, including emission-line placement, photometric depth, luminosity, and temporal baseline.

\begin{table*}
    \centering
    \caption{Number of selected CLAGN candidates for different $\chi^2_{r,\rm narrow}$ thresholds. Recovery rate of \GW{} refers to the 73 overlapping sources. Median $|\Delta g|$, frac($|\Delta g|>1$), and median
$\Delta\log L_{\rm cont}$ are computed on the DESI-DR1 comparison only.}
    \label{tab:chi2_thresholds}
    \begin{tabular}{lcccccc}
        \hline
        Threshold & $N_{\rm obs}$ & $N_{\rm unique}$ & GW25 recovery & median $|\Delta g|$ & frac ($|\Delta g|>1$)& median $\Delta \text{log L}_\text{cont}$\\
        \hline
        $\chi^2_{r,\rm narrow} > 5$  & 23,797 & 17,399 & 68.5\% &0.51 &7.7\% &0.20 dex \\
        $\chi^2_{r,\rm narrow} > 10$ & 8,512  & 6,469  & 49.3\% & 0.52 & 8.2\% & 0.21 dex \\
        $\chi^2_{r,\rm narrow} > 15$ & \obscand{}  & \CAND{}  & 41.1\% & 0.58 & 10.3\% & 0.23 dex\\
        $\chi^2_{r,\rm narrow} > 25$ & 1,553  & 1,261  & 30.1\% & 0.64 & 14.2\% & 0.26 dex\\
        \hline
        
    \end{tabular}
    \tablecomments{Median $|\Delta g|$ and frac($|\Delta g|>1$) are computed from the
DESI-DR1 comparison for each object (the primary comparison used in candidate
selection, Sect.~3.7). This differs
negligibly from the frac($|\Delta g|>1$)$=14\%$ reported in Sect.~4.4
at $\chi^2_{r,\rm narrow}>15$, which also considers SDSS/BOSS comparison.
$\Delta\log L_{\rm cont} = 0.4\,|\Delta g|$ is the implied fractional change in
continuum flux under a pure photometric amplitude interpretation and does not
correct for bolometric or aperture effects.}
\end{table*}

\subsection{Comparison with spectroscopically-selected CLAGN}

\begin{figure*}[!ht]
\centering
\includegraphics[width=\textwidth]
{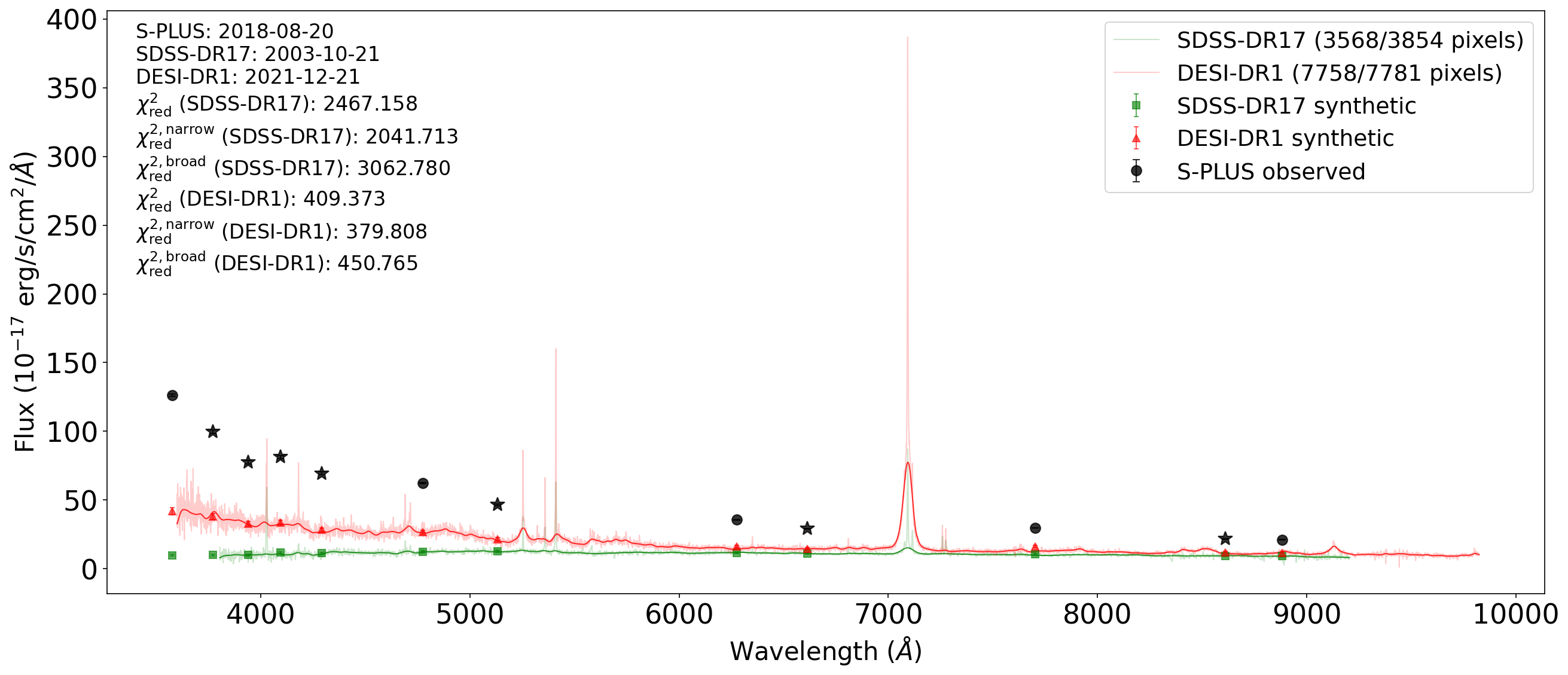}
  \caption{Quasar observed in 2000 by SDSS-DR17, 2018 by S-PLUS and 2021 by DESI-DR1. S-PLUS broad and narrow bands are shown in points and star markers, respectively. For visualization purposes only, the spectra were cleaned using iterative $3\sigma$ clipping to remove outlying pixels, followed by interpolation over masked regions and Gaussian smoothing to reduce pixel-scale noise.} 
     \label{fig:example_qso3}
\end{figure*}
Figure \ref{fig:example_qso3} illustrates an example comparison between SDSS-DR17 (observed on 2003-10-21), DESI-DR1 (observed on 2021-12-21), and S-PLUS observations for a highly variable quasar candidate. The spectra exhibit substantial changes in continuum level and emission-line strength, particularly around [O III] $\lambda = 5007$ and H$\alpha$ $\lambda = 6563$. The large reduced chi-squared values, particularly for SDSS-DR17 ($\chi^2_r = 2360.594$), further quantify the poor agreement between the spectral flux and the S-PLUS photometry at the earlier epoch, corroborating the classification of this source as a strong CLAGN candidate selected by our pipeline. Notably, the DESI-DR1 spectrum yields a significantly lower $\chi^2_r$ (521.840), yet still deviates from the S-PLUS photometry, suggesting that residual variability or calibration offsets may persist even in the more recent observation. Because our uncertainty model does not include the full covariance structure of spectrophotometric calibration and band-to-band correlations, $\chi^2_r$ should not be interpreted as a formal goodness-of-fit statistic. Instead, it serves as an empirical variability indicator.
\begin{figure}[!ht]
\centering
\includegraphics[width=\columnwidth]
{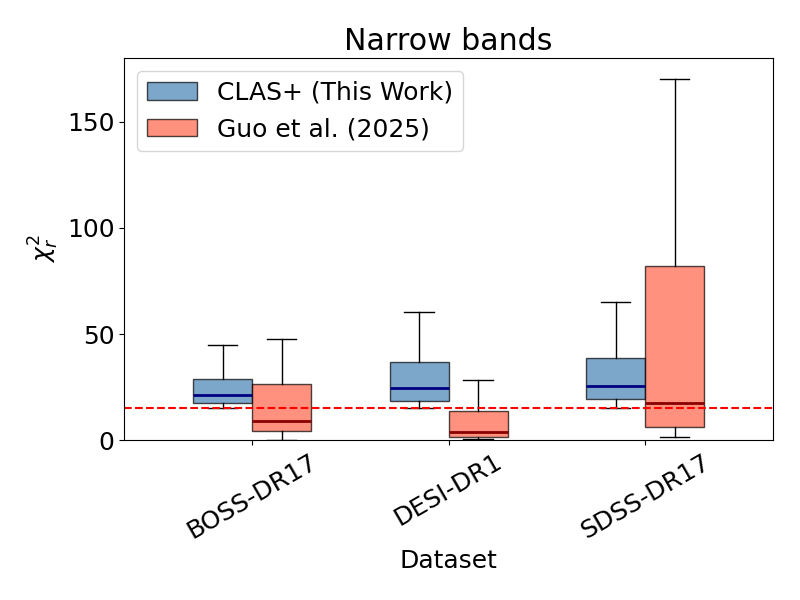}
  \caption{Distribution of $\chi^2_{r}$ computed between the S-PLUS photometric fluxes and the synthetic fluxes convolved from the spectra through the seven narrow-band filters for the CLAS+ catalog and \GW{}. The red dashed line indicates the selection threshold $\chi^2_{r} > 15$ adopted in this work to identify candidates with significant photometric excess. Outliers are excluded for clarity.}
     \label{fig:dist}
\end{figure}

\citealt[hereafter \GW,]{2025ApJS..278...28G} published a catalog of 561 CLAGN from DESI DR1 and SDSS DR16, of which 139 are in $\text{Dec}<2\,\text{deg}$ (approximately the maximum declination of S-PLUS' footprint). We cross-match this catalog with S-PLUS DR6, finding a total of 85 sources in overlapped regions, of which 12 were excluded by the emission-line masking process.
{Figure \ref{fig:dist}} shows the distribution of $\chi^2_r$ for the 73 sources in the \GW{} sample and for all CLAGN candidates selected in this work. This comparison shows that a subset of spectroscopically selected GW25 sources also displays significant narrow-band photometric variability according to our metric. The limited overlap with \GW{} is expected because the two catalogs probe different variability regimes. While \GW{} identifies spectroscopic changing-look transitions, CLAS+ selects the high-variability tail of the S-PLUS narrow-band $\chi^2_r$ distribution. As shown in Fig.~\ref{fig:dist}, most \GW{} sources have lower narrow-band variability according to our metric, whereas CLAS+ is intentionally optimized for high-confidence, high-amplitude candidates. At the adopted threshold, 41.1\% of the 73 GW25 sources pass the $\chi^2_{r,\rm narrow}>15$ criterion, while 14 sources satisfy all final CLAS+ selection criteria and are included in both catalogs. We also note that most of the CLAGN literature focuses on northern hemisphere surveys, resulting in limited overlap with our sample. Nevertheless, upcoming southern hemisphere spectroscopic surveys, such as 4-metre Multi-Object Spectroscopic Telescope (4MOST; \citealt{2019Msngr.175....3D}), will provide multi-epoch spectroscopy over a large area of the southern sky, further highlighting the value of a pipeline combining S-PLUS photometry with spectroscopic data to identify and characterize changing-look events in the southern hemisphere.

As summarized in Table~\ref{tab:clagn_systematic}, CLAS+ substantially increases the number of high-priority CLAGN candidates identified through systematic searches. If confirmed through spectroscopic follow-up, these candidates would significantly expand the known CLAGN population, highlighting the potential of narrow-band photometry for identifying changing-look candidates at large scale.

\begin{table*}
\centering
\caption{Systematic searches for changing-look AGN.}
\label{tab:clagn_systematic}
\footnotesize
\setlength{\tabcolsep}{3pt}
\begin{tabular*}{\textwidth}{@{\extracolsep{\fill}} l l c c c l}
\toprule
Data & Selection Method & $z$ range & $N_{candidates}$ & $N_{new}$ & Reference \\
\midrule
SDSS DR7, BOSS, PS1 & Broad-band variability & $0.20<z<0.63$ & 1,011 &10 & \citet{2016MNRAS.457..389M} \\
SDSS DR14, LAMOST DR5 & Broad-band variability and spectral transitions & $0.08<z<0.58$ & 10,263 &21 & {\citet{2018ApJ...862..109Y}} \\
ZTF, SDSS DR12 & Light curve & $z<0.17$ & -- &6 & {\citet{2019ApJ...883...31F}} \\
SDSS DR7, PS1, CRTS & Broad-band variability& $z<0.83$ & 262 &17 &{\citet{2019ApJ...874....8M}}\\
SDSS MaNGA, SDSS DR7, BOSS & Catalog match and spectral transitions & $0.026<z<0.107$ & 9 &4 &{\citet{2020MNRAS.497..192H}}\\
SDSS DR7 & Spectral transitions & $0.1<z<0.3$ & 941 & 4 & \citet{2021AA...650A..33P} \\
SkyMapper DR1, PS1 & Broad-band Variability & $z<0.04$ & 127 &2 & \citet{2021MNRAS.503.2583S} \\
SkyMapper DR3, 6dFGS & Broad-band Variability & $z<0.1$ & 157 &25 & \citet{2022MNRAS.511...54H} \\
SDSS-IV & Spectral transitions & $z<0.9$ & 61 & 15 & \citet{2022ApJ...933..180G} \\
ZTF DR15, SDSS DR14 & Light curve & $z<0.35$ & 47 &21 &{\citet{2024ApJ...966..128W}} \\
DESI EDR, SDSS DR16 & Spectral transitions & -- & 221 &56 & \citet{2024ApJS..270...26G} \\
ATLAS, 6dFGS & Light curve & $z<0.1$ & 201 &51 & \citet{2024MNRAS.535.2322A} \\
SDSS-V, SDSS DR16 & Spectral transitions & $0.06<z<2.4$ & 3,338 &107 & \citet{2024ApJ...966...85Z}    \\
DESI DR1, SDSS DR16 & Variability & $z<0.9$ & 924 &561 &  \citet{2025ApJS..278...28G} \\
S-PLUS DR6, DESI DR1, SDSS DR17 & Narrow-band variability & $z<4.71$ & \CAND{} &\newcand{}$^*$ & CLAS+ (this work) \\
\bottomrule
\end{tabular*}
\tablecomments{$^\ast$Number of previously unreported CLAS+ candidates; spectroscopic follow-up is required for confirmation.
Acronyms: ATLAS = Asteroid Terrestrial-impact Last Alert System;
BOSS = Baryon Oscillation Spectroscopic Survey;
CRTS = Catalina Real-Time Transient Survey;
DESI = Dark Energy Spectroscopic Instrument;
DR = Data Release;
EDR = Early Data Release;
LAMOST = Large Sky Area Multi-Object Fiber Spectroscopic Telescope;
MaNGA = Mapping Nearby Galaxies at Apache Point Observatory;
PS1 = Pan-STARRS1;
SDSS = Sloan Digital Sky Survey;
S-PLUS = Southern Photometric Local Universe Survey;
ZTF = Zwicky Transient Facility;
6dFGS = Six-degree Field Galaxy Survey.
}
\end{table*}

\subsection{Photometric turn-on, turn-off, or intermediate stage}
\label{sec:classification}

Sources showing only a brightening-up transition were classified as photometric turn-on, those showing only a fading-down transition as photometric turn-off, and those exhibiting both signatures across epochs were classified as photometric intermediate (see Section \ref{ssec:turn}). In this context, the intermediate class represents objects with mixed or transitional behavior that cannot be described by a single monotonic state change. Note that our classification does not adopt the standard definition of turn-on (appearance of broad emission lines) and turn-off (their disappearance) commonly used in the literature \citep[e.g.][]{2024ApJS..272...13P}, as spectroscopic follow-up would be needed for this evaluation.


Among the CLAS+ spectroscopic--photometric comparisons with reliable temporal classification, 40\% are classified as photometric turn-on, 41\% as intermediate, and 19\% as photometric turn-off. In Figure ~\ref{fig:redshift_state}, we present the redshift evolution of the fractions of turn-on, turn-off, and intermediate states. At low redshift, the relative fractions of the three classes are comparable. However, at $z \approx 1.2$, the turn-on population becomes dominant, accounting for more than 50\% of the sources, while the turn-off fraction drops to below 20\%, suggesting a redshift-dependent asymmetry in the observed CLAS+ photometric classifications. The highest-redshift bin should be interpreted with particular caution because it is more sensitive to small-number statistics and redshift-dependent selection effects.

\subsection{Redshift-dependent selection effects}

The redshift dependence shown in Fig.~\ref{fig:redshift_state} should be interpreted in the context of the CLAS+ selection function. Our method is sensitive only to redshift intervals in which at least one strong quasar emission line falls within an S-PLUS narrow-band filter. Therefore, the probability of detecting a photometric discrepancy is not uniform with redshift, but depends on the line--filter combination, filter throughput, line equivalent width, source luminosity, photometric depth, and the time baseline between the spectroscopic and S-PLUS epochs. Cosmological time dilation further reduces the rest-frame interval sampled by a fixed observer-frame baseline, so current survey lengths may not capture the full temporal evolution of CLAGN transitions at higher redshifts. These effects can bias the observed state fractions by preferentially selecting events with particular amplitudes, timescales, emission lines, or phases of variability. As a result, the observed fractions of photometric turn-on, turn-off, and intermediate candidates do not represent completeness-corrected intrinsic CLAGN rates. Instead, they describe the relative state fractions within the subset of quasars for which CLAS+ is sensitive to narrow-band emission-line variability. We therefore do not interpret the redshift trend as direct evidence for intrinsic AGN evolution. A physical interpretation would require completeness modeling or injection--recovery simulations as a function of redshift, emission line, variability amplitude, source luminosity, and temporal baseline.

We verified that the turn-on/turn-off/intermediate fractions shown in Fig.~\ref{fig:redshift_state} are stable under moderate changes to the $\chi^2_\text{r}$ threshold as the turn-on fraction remains within 46–47\% for thresholds between 10 and 30. The redshift distribution (Fig.~\ref{fig:redshift}) is statistically indistinguishable between $\chi^2_\text{r}>15$ and $\chi^2_\text{r}>20$ (KS test, p=0.38) but differs significantly for larger changes to the threshold ($\chi^2_\text{r}>25$ or $>30$, $p<10^{-4}$), indicating that substantially tighter thresholds preferentially select a modestly lower-redshift subsample. This should be kept in mind when comparing our redshift-dependent state fractions across studies using different variability thresholds.

\subsection{Broad-band variability selection}
Previous photometric searches for CLAGN have often targeted large-amplitude continuum variability. For example, \cite{2016MNRAS.457..389M} adopted $\,|\,\Delta g\,|\,>1$ mag, while \cite{2019ApJ...874....8M} required $\,|\,\Delta g\,|\,>1$ mag and $\,|\,\Delta r\,|\,>0.5$ mag. More recently, \cite{2026ApJ..1004..121V} proposed the less restrictive criterion $\,|\,\Delta g\,|\,>0.4$ mag together with $\,|\,\Delta(g-r)\,|\,>0.2$ mag. To place CLAS+ in this context, we measured the broad-band variability between the spectroscopic synthetic photometry and the S-PLUS epoch. The median absolute g-band variation is 0.58 mag, with 75.9\% of the CLAS+ comparisons satisfying $\,|\,\Delta g\,|\,>0.4$ mag and 14\% satisfying $\,|\,\Delta g\,|\,>1$ mag. Only 13.7\% simultaneously satisfy the MacLeod et al. (2019) criteria of $\,|\,\Delta g\,|\,>1$ mag and $\,|\,\Delta r\,|\,>0.5$ mag.

Thus, although CLAS+ preferentially selects strongly variable systems, a substantial fraction of the candidates would not be recovered by the classical $\,|\,\Delta g\,|\,>1$ mag criterion. This difference reflects the use of narrow-band information in CLAS+, which enhances sensitivity to variability associated with strong emission-line regions even when the corresponding broad-band continuum variation is more moderate. Indeed, the S-PLUS narrow bands exhibit a more extended high-amplitude tail than the broad bands (Fig. \ref{fig:delta}).

As $\chi^2_\text{r}$ increases from 5 to 25, the median $|\Delta g|$ of selected sources rises from 0.51 to 0.64 mag, and the fraction satisfying the \cite{2016MNRAS.457..389M} $|\Delta g|>1$ mag criterion rises from 7.7\% to 14.2\% (Table \ref{tab:chi2_thresholds}). Matching $\chi^2_\text{r}$ to the \cite{2016MNRAS.457..389M} $|\Delta g|>1$ mag criterion by source count alone would require $\chi^2_\text{r} \geq 35–40$, well above our adopted threshold, underscoring that CLAS+ and broad-band amplitude selection are sensitive to different (correlated but distinct) variability signatures rather than two thresholds on the same underlying quantity. Converting $|\Delta g|$ to an implied continuum flux ratio ($\Delta \text{log L} = 0.4|\Delta g|$), the median continuum luminosity change rises from 0.20 dex (factor 1.6$\times$) at $\chi^2_\text{r}>5$ to 0.26 dex (factor 1.8$\times$) at $\chi^2_\text{r}>25$. This is systematically smaller than the median $\Delta \text{log L}_{5100} \approx 0.43$ dex (dim: log L$_{5100}=43.67$; bright: log L$_{5100}=44.10$ erg s$^{-1}$) reported for spectroscopically confirmed CLAGN by GW25.  We interpret this as consistent with CLAS+ recovering a population with genuine but comparatively modest continuum-luminosity variability, selected primarily via line-region response rather than large-amplitude continuum flares, i.e., a population plausibly under-represented in spectroscopically pre-selected or broad-band-amplitude-selected CLAGN samples.

\begin{figure*}[!ht]
\centering
\begin{subfigure}{\textwidth}
    \centering
    \includegraphics[width=0.9\linewidth]{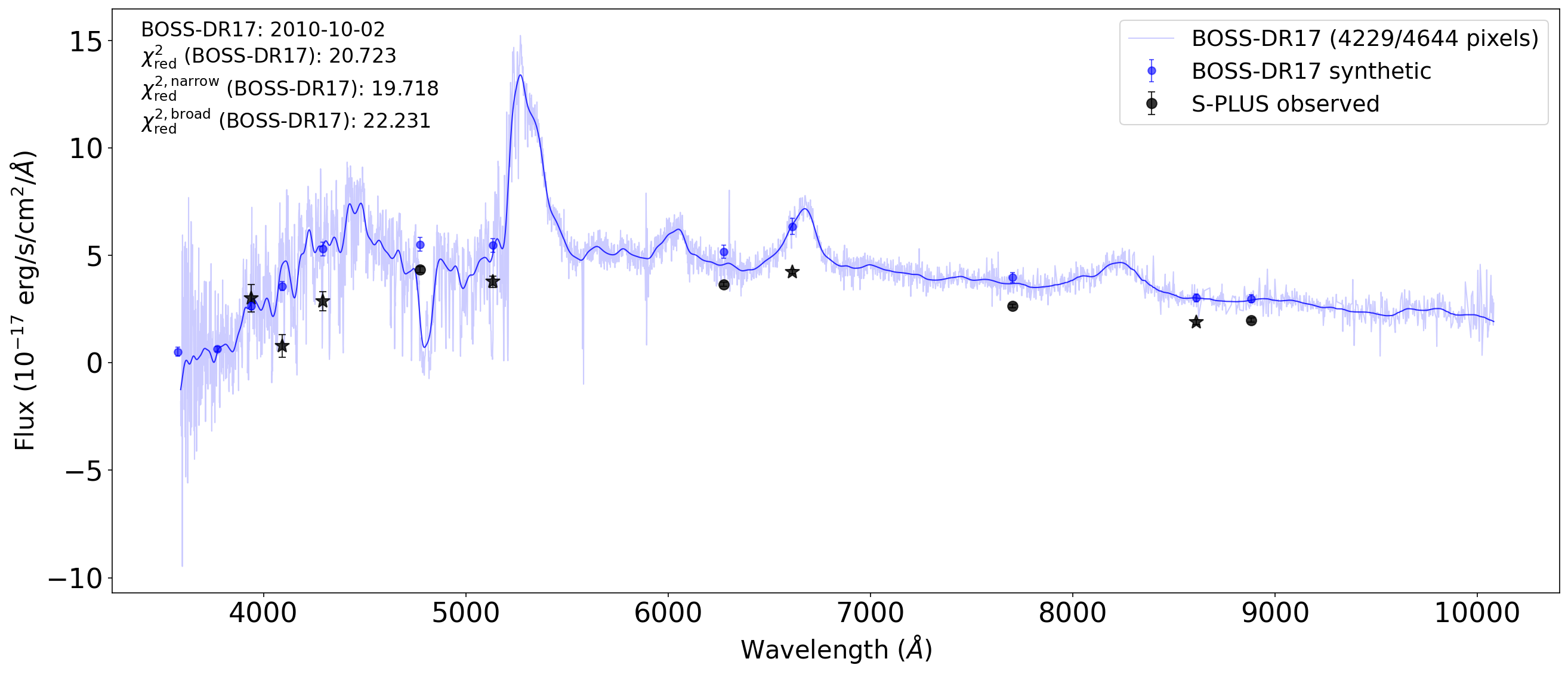}
    \caption{Turn-off CLAGN at redshift 3.32. The S-PLUS observation is from 2018-10-07.}
    \label{fig:turnoff}
\end{subfigure}

\vspace{0.3cm}

\begin{subfigure}{\textwidth}
    \centering
    \includegraphics[width=0.9\linewidth]{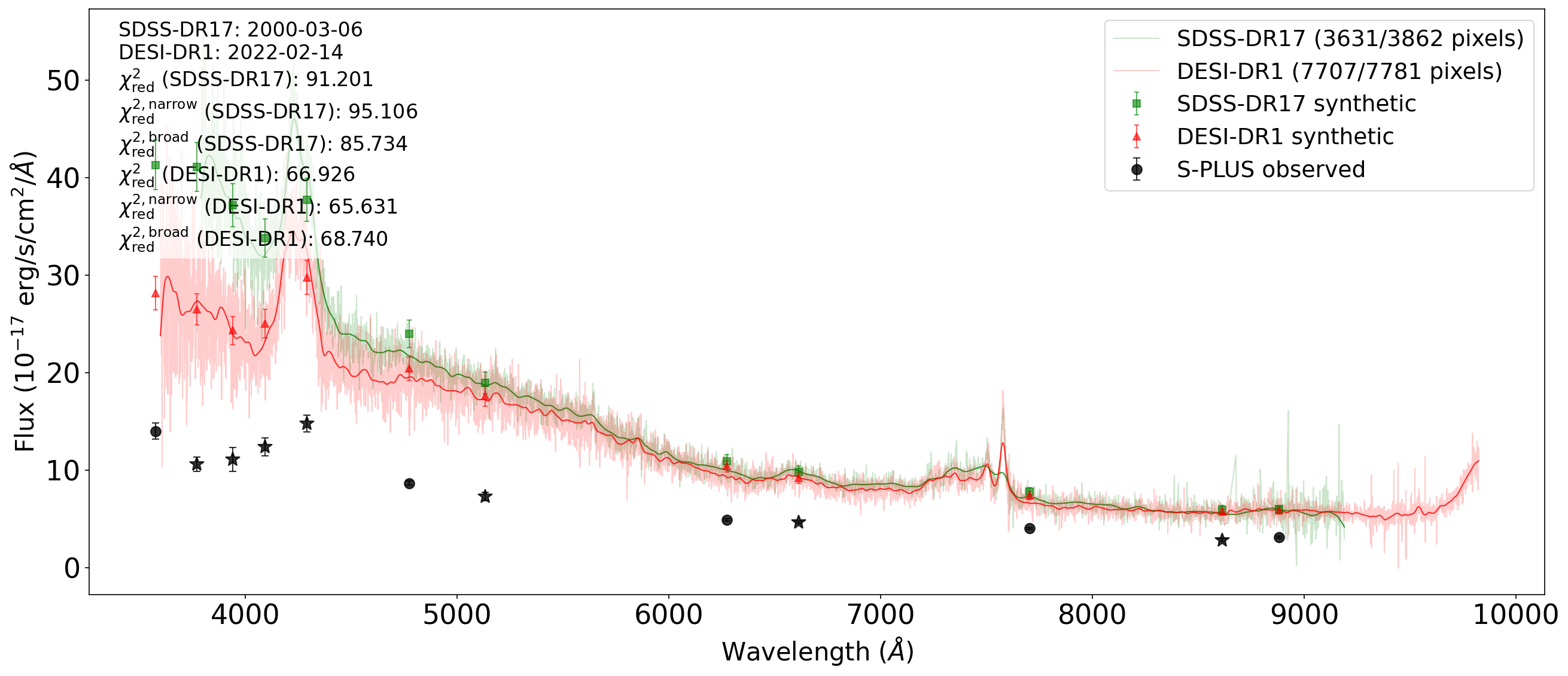}
    \caption{Intermediate CLAGN at redshift 0.5136. The S-PLUS observation is from 2021-04-14.}
    \label{fig:interm}
\end{subfigure}

\begin{subfigure}{\textwidth}
    \centering
    \includegraphics[width=0.9\linewidth]{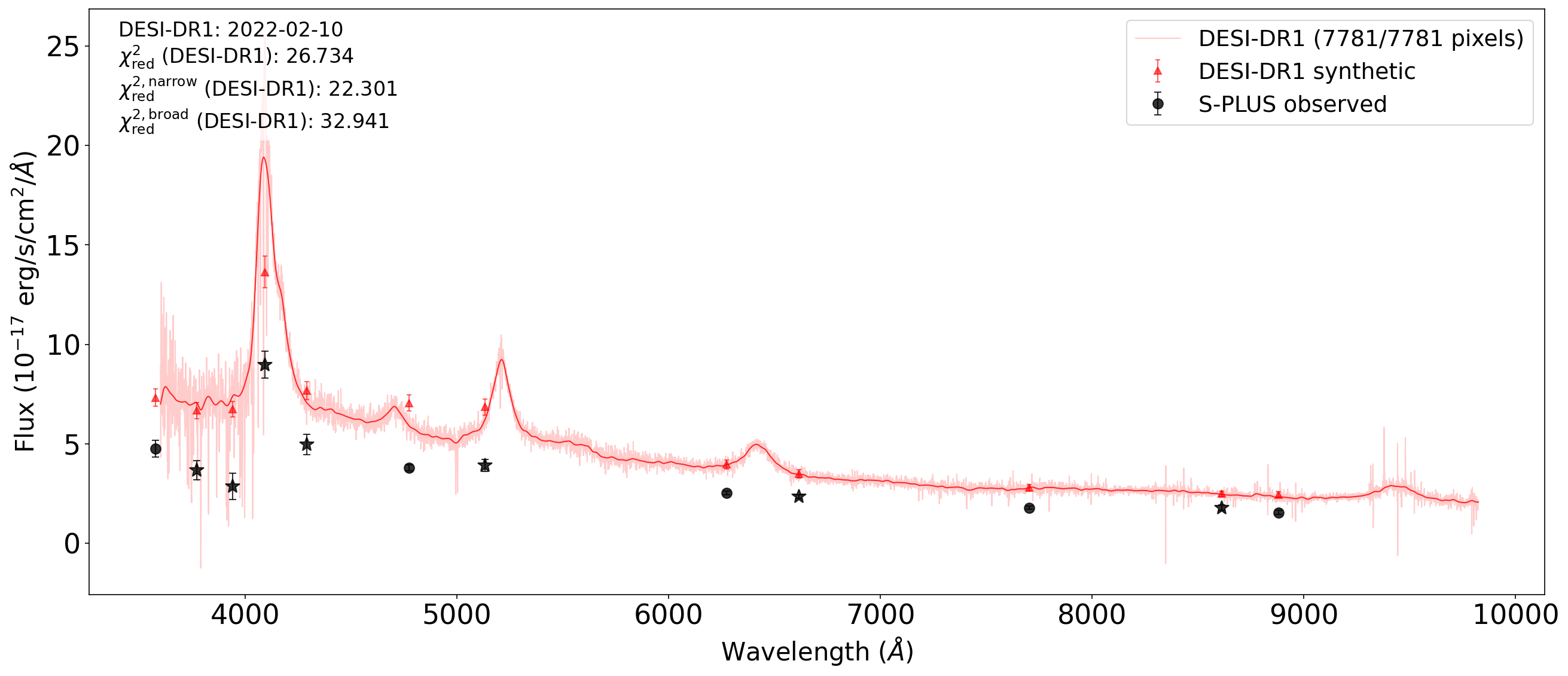}
    \caption{Turn-on CLAGN at redshift 2.3639. The S-PLUS observation is from 2019-03-05.}
    \label{fig:turnon}
\end{subfigure}
\caption{Examples of a turn-on, intermediate, and turn-off CLAGN candidates. }
\label{fig:examples}
\end{figure*}

\section{Description of the CLAS+ catalog}
\label{sec:catalog}
\begin{figure*}[!t]
    \centering
    \includegraphics[width=\linewidth]{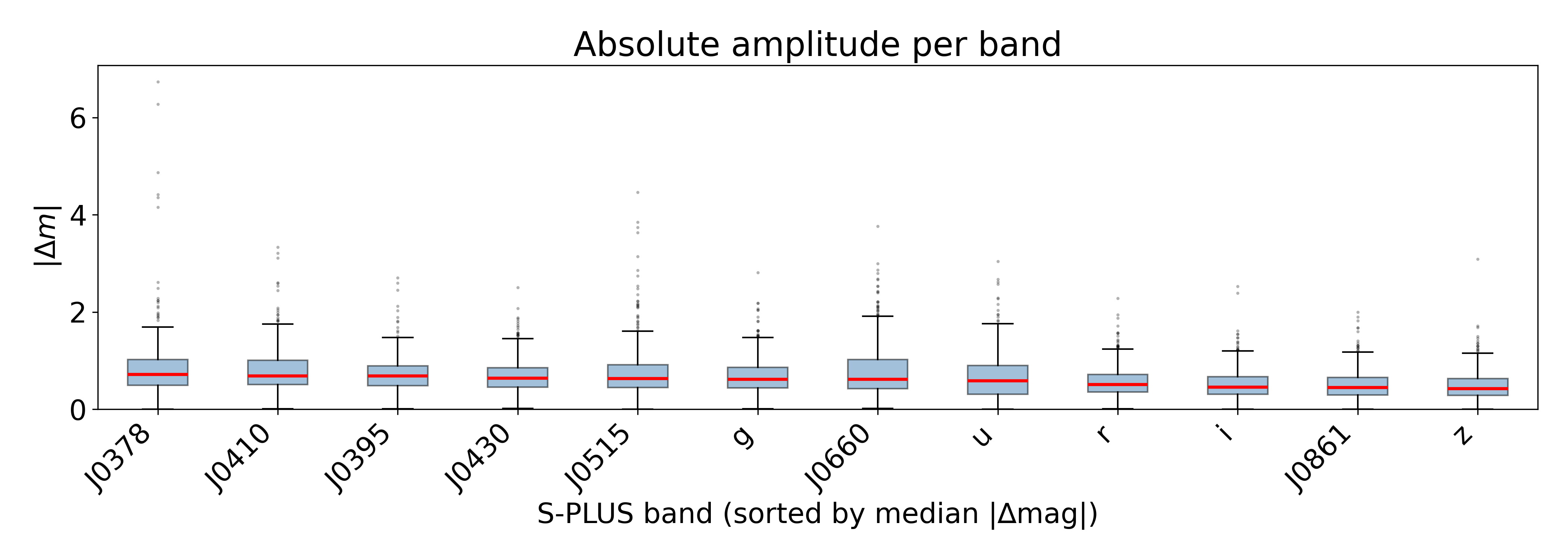}
\caption{Distribution of the absolute variability amplitude across S-PLUS filters for the CLAS+ catalog, ordered by decreasing median amplitude from left to right. Bluer filters exhibit a systematically higher variability than redder ones. Notably, the narrow-band filters capture a greater number of large $|\Delta m|$ values compared to the broad-band filters, as highlighted by the presence of pronounced outliers.}
\label{fig:delta}
\end{figure*}

Our final catalog contains \obscand{} observations of \CAND{} CLAGN candidates, representing 3.6\% out of 98,139 quasars in DESI DR1. For the spectroscopic--photometric comparisons with reliable temporal classification, we classify 40\% as photometric turn-on, 41\% as intermediate, and 19\% as photometric turn-off. The intermediate class is defined by transitional behavior inferred from the relative timing of the S-PLUS photometry with respect to the DESI and SDSS/BOSS spectroscopy, whenever available. Examples of a turn-on source at $z = 2.36$, a turn-off source at $z = 3.32$, and an intermediate-state source at $z = 0.51$ are shown in Fig.~\ref{fig:examples}. In particular, Figs.~\ref{fig:turnoff} and \ref{fig:interm} display a clear variability of prominent emission lines, traced by the J0430 and J0410 filters, respectively. 

The catalog spans a broad range of redshifts, enabling the study of changing-look phenomena across different AGN regimes. We also provide key observational properties for each source, including variability metrics, photometric measurements, and spectroscopic information, allowing flexible selection and follow-up analyses. The distribution of absolute variability amplitude ($|\Delta m|$) across filters reveals a clear wavelength dependence, with bluer filters showing systematically higher variability than redder ones (Fig. \ref{fig:delta}). In addition, narrow-band filters tend to capture more extreme $|\Delta m|$ values than broad-band filters, as indicated by the presence of pronounced outliers. The emission-line constraint described in Section \ref{ssec:mask} ensures that the observed narrow-band excess is plausibly associated with variability in a known quasar emission line, strengthening the physical interpretation of the candidates. However, it also introduces a redshift-dependent selection function, since different emission lines enter the S-PLUS narrow-band filters at different redshifts. As a result, CLAS+ should be interpreted as a high-confidence candidate catalog rather than a complete census of CLAGN.

The full catalog, including the unselected sources, together with the software used to generate the catalog, is publicly available at \cite{nakazono_2026_21843309}. A detailed description of all catalog columns is given in Table~\ref{tab:description}.

\begin{table*}[ht]
\centering
\caption{Description of columns in the final CLAS+ catalog.}
\label{tab:clas_columns}
\begin{tabular}{llll}
\toprule
\textbf{Column} & \textbf{Type} & \textbf{Unit} & \textbf{Description} \\
\midrule
splus\_id & string & -- & S-PLUS object identifier. \\
ra & float & deg & Right ascension (J2000). \\
dec & float & deg & Declination (J2000). \\
redshift & float & -- & Spectroscopic redshift. \\
dataset & string & -- & Spectroscopic dataset used in the comparison (e.g., DESI-DR1). \\
spectrum\_obs\_date & date/datetime & UTC date & Observation date of the spectrum used to compute synthetic photometry. \\
spectrum\_obs\_date\_mjd & float & -- & MJD of the spectrum used to compute synthetic photometry. \\
chi2\_red & float & -- & Reduced chi-square over all valid filters, $\chi^2_{\rm red}=\chi^2/\mathrm{dof}$. \\
dof & int & -- & Number of valid filters used in \texttt{chi2}. \\
chi2\_red\_broad & float & -- & Reduced chi-square for broad bands, $\chi^2_{\rm broad}/\mathrm{dof}_{\rm broad}$. \\
dof\_broad & int & -- & Number of valid broad bands used in \texttt{chi2\_broad}. \\
chi2\_red\_narrow & float & -- & Reduced chi-square for narrow bands, $\chi^2_{\rm narrow}/\mathrm{dof}_{\rm narrow}$. \\
dof\_narrow & int & -- & Number of valid narrow bands used in \texttt{chi2\_narrow}. \\
state\_consolidated & string & -- & Per-epoch temporal state (e.g., turn-on, turn-off, intermediate). \\
temporal\_distance\_days & int & days & Number of days between observed states \\
n\_unique\_spectra & int & -- & Number of valid spectra for this source. \\
mag\_psf\_[filter] & float & mag & S-PLUS magnitude for each [filter] \\
err\_mag\_psf\_[filter] & float & mag & S-PLUS error magnitude for each [filter] \\
flags\_det & int & -- & S-PLUS photometric quality flag in the detection image \\
splus\_obs\_date & date/datetime & UTC date & Observation date of the S-PLUS observation. \\
splus\_obs\_date\_mjd & float & -- & MJD of the S-PLUS observation. \\
\bottomrule
\label{tab:description}
\end{tabular}
\end{table*}

\section{Conclusions and future work}
\label{sec:conclusions}
In this work, we presented CLAS+, a catalog of changing-look AGN candidates selected by comparing spectroscopically derived synthetic photometry with S-PLUS broad- and narrow-band observations. This approach combines the spectral information provided by surveys such as SDSS and DESI with the wide-field narrow-band photometry of S-PLUS, enabling an efficient search for strong photometric--spectroscopic discrepancies associated with AGN variability.

Applying our pipeline to 98,139 quasars in the S-PLUS footprint, we identify a high-confidence catalog of \CAND{} CLAGN candidates. A crossmatch with the spectroscopically selected CLAGN catalog of GW25 shows that our method recovers a subset of previously known sources, providing an external consistency check of the technique.

We classify the selected spectroscopic--photometric comparisons into photometric turn-on, photometric turn-off, and intermediate cases based on the sign and temporal behavior of the narrow-band residuals. Among the classified comparisons, the sample is composed of photometric turn-on (\turnon{}), intermediate (\intermediate{}), and photometric turn-off (\turnoff{}) cases. Intermediate sources are particularly useful targets for follow-up because they may trace non-monotonic or complex variability behavior across multiple epochs.

The classified sample shows an apparent redshift-dependent change in the relative state fractions: around $z\approx1.2$, photometric turn-on candidates become more common, while the photometric turn-off fraction decreases. However, this trend should be regarded as an observed property of the CLAS+ selected sample, rather than a direct measurement of intrinsic CLAGN evolution. The observed fractions are affected by the CLAS+ selection function, including emission-line placement within the S-PLUS filters, rest-frame wavelength coverage, filter throughput, photometric depth, source luminosity, and the time baseline between spectroscopic and photometric observations. A physical interpretation of the redshift trend would require completeness modeling or injection--recovery simulations as a function of redshift, emission line, variability amplitude, luminosity, and temporal baseline.

Future work will focus on dedicated spectroscopic follow-up to confirm and characterize the CLAS+ candidates. We will also expand the methodology to complementary narrow-band surveys, such as J-PAS \citep{2014arXiv1403.5237B}, and scale the pipeline to upcoming large time-domain surveys such as the Legacy Survey of Space and Time (LSST; \citealt{2019ApJ...873..111I}). These efforts will improve the confirmation rate, quantify the selection function, and enable more robust population studies of changing-look AGN.


\begin{acknowledgments}

L. N. acknowledges FAPESP grant number 2024/07281-0. R. R. V. acknowledges the support from CAPES (grant 88887.821818/2023-00) and FAPESP (grants 2023/05003-0 and 2024/16592-9). I would like to thank Julia Thainá da Silva Cunha Batista and the referee for very useful comments. S. P. is supported by the international Gemini Observatory, a program of NSF NOIRLab, which is managed by the Association of Universities for Research in Astronomy (AURA) under a cooperative agreement with the U.S. National Science Foundation, on behalf of the Gemini partnership of Argentina, Brazil, Canada, Chile, the Republic of Korea, and the United States of America. R.D. gratefully acknowledges support by the ANID BASAL project FB210003

The S-PLUS project, including the T80-South robotic telescope and the S-PLUS scientific survey, was founded as a partnership between the Fundação de Amparo à Pesquisa do Estado de São Paulo (FAPESP), the Observatório Nacional (ON), the Federal University of Sergipe (UFS), and the Federal University of Santa Catarina (UFSC), with important financial and practical contributions from other collaborating institutes in Brazil, Chile (Universidad de La Serena), and Spain (Centro de Estudios de Física del Cosmos de Aragón, CEFCA). We further acknowledge financial support from the São Paulo Research Foundation (FAPESP) grant 2019/263492-3, the Brazilian National Research Council (CNPq), the Coordination for the Improvement of Higher Education Personnel (CAPES), the Carlos Chagas Filho Rio de Janeiro State Research Foundation (FAPERJ), and the Brazilian Innovation Agency (FINEP). The members
of the S-PLUS collaboration are grateful for the contributions from CTIO staff in helping in the construction, commissioning and maintenance of the T80-South
telescope and camera. 
This research uses services or data provided by the SPectra Analysis and Retrievable Catalog Lab (SPARCL), which is part of the Community Science and Data Center (CSDC) program at NSF National Optical-Infrared Astronomy Research Laboratory. NOIRLab is operated by the Association of Universities for Research in Astronomy (AURA), Inc. under a cooperative agreement with the National Science Foundation.
Funding for the SDSS and SDSS-II has been provided by the Alfred P. Sloan Foundation, the Participating Institutions, the National Science Foundation, the U.S. Department of Energy, the National Aeronautics and Space Administration, the Japanese Monbukagakusho, the Max Planck Society, and the Higher Education Funding Council for England. The SDSS website is http://www.sdss.org/.

The SDSS is managed by the Astrophysical Research Consortium for the Participating Institutions. The Participating Institutions are the American Museum of Natural History, Astrophysical Institute Potsdam, University of Basel, University of Cambridge, Case Western Reserve University, University of Chicago, Drexel University, Fermilab, the Institute for Advanced Study, the Japan Participation Group, Johns Hopkins University, the Joint Institute for Nuclear Astrophysics, the Kavli Institute for Particle Astrophysics and Cosmology, the Korean Scientist Group, the Chinese Academy of Sciences (LAMOST), Los Alamos National Laboratory, the Max-Planck-Institute for Astronomy (MPIA), the Max-Planck-Institute for Astrophysics (MPA), New Mexico State University, Ohio State University, University of Pittsburgh, University of Portsmouth, Princeton University, the United States Naval Observatory, and the University of Washington.
\end{acknowledgments}

\bibliography{ApJ}{}
\bibliographystyle{aasjournalv7}



\end{document}